\documentclass[default,iicol]{sn-jnl}

\usepackage{graphicx}%
\usepackage{multirow}%
\usepackage{multicol}
\usepackage{amsmath,amssymb,amsfonts}%
\usepackage{amsthm}%
\usepackage{mathrsfs}%
\usepackage[title]{appendix}%
\usepackage{xcolor}%
\usepackage{textcomp}%
\usepackage{manyfoot}%
\usepackage{booktabs}%
\usepackage{algorithm}%
\usepackage{algorithmicx}%
\usepackage{algpseudocode}%
\usepackage{listings}%

\usepackage{ragged2e}
\usepackage{balance}
\usepackage{multicol}
\usepackage[misc,geometry]{ifsym}
\usepackage{1-in-2}

\begin{document}

\title{Cryogenic quasi-static embedded DRAM for energy-efficient compute-in-memory applications}

\author{
        Yuhao Shu\textsuperscript{1,2,3$\dagger$},
		Hongtu Zhang\textsuperscript{1,2,3$\dagger$},
		Hao Sun\textsuperscript{1,2,3},
        Mengru Zhang\textsuperscript{4,5},
        Wenfeng Zhao\textsuperscript{6},
        Qi Deng\textsuperscript{1,3,7},
        Zhidong Tang\textsuperscript{1},
        Yumeng Yuan\textsuperscript{1,2,3},
        Yongqi Hu\textsuperscript{1,2,3},
        Yu Gu\textsuperscript{1,2,3},
        Xufeng Kou\textsuperscript{1,8}\Letter,
        Yajun Ha\textsuperscript{1,8}\Letter
}


\abstract{
Compute-in-memory (CIM) presents an attractive approach for energy-efficient computing in data-intensive applications.
However, the development of suitable memory designs to achieve high-performance CIM remains a challenging task.
Here, we propose a cryogenic quasi-static embedded DRAM to address the logic-memory mismatch of CIM.
Guided by the re-calibrated cryogenic device model, the designed four-transistor bit-cell achieves full-swing data storage, low power consumption, and extended retention time at cryogenic temperatures. 
Combined with the adoption of cryogenic write bitline biasing technique and readout circuitry optimization, our 4Kb cryogenic eDRAM chip demonstrates a 1.37$\times$10$^6$ times improvement in retention time, while achieving a 75 times improvement in retention variability, compared to room-temperature operation. 
Moreover, it also achieves outstanding power performance with a retention power of 112 fW and a dynamic power of 108 $\mu$W at 4.2 K, which can be further decreased by 7.1\% and 13.6\% using the dynamic voltage scaling technique.
This work reveals the great potential of cryogenic CMOS for high-density data storage and lays a solid foundation for energy-efficient CIM implementations.
\vspace{-2em}
}

\maketitle

\mathfootnote{
\footnotesize
\thanks{$^{\rm{1}}$School of Information Science and Technology, ShanghaiTech University, Shanghai, China.}
\thanks{$^{\rm{2}}$Shanghai Institute of Microsystem and Information Technology, Chinese Academy of Sciences, Shanghai, China.}
\thanks{$^{\rm{3}}$School of Electronic, Electrical and Communication Engineering, University of Chinese Academy of Sciences, Beijing, China.}
\thanks{$^{\rm{4}}$State Key Laboratory of Integrated Chip and System, Frontier Institute of Chip and System, Fudan University, Shanghai, China.}
\thanks{$^{\rm{5}}$State Key Laboratory of ASIC and System, School of Microelectronics, Fudan University, Shanghai, China.}
\thanks{$^{\rm{6}}$Department of Electrical and Computer Engineering, Binghamton University SUNY, Binghamton, NY, USA.}
\thanks{$^{\rm{7}}$Shanghai Advanced Research Institute, Chinese Academy of Sciences, Shanghai, China.}
\thanks{$^{\rm{8}}$Shanghai Engineering Research Center of Energy Efficient and Custom AI IC, Shanghai, China. $^\dagger$These authors contributed equally to this work. \Letter~e-mail: \href{mailto:kouxf@shanghaitech.edu.cn}{kouxf$@$shanghaitech.edu.cn}; \href{mailto:hayj@shanghaitech.edu.cn}{hayj$@$shanghaitech.edu.cn}}
}

\begin{figure*}[htp]
    \centerline{\includegraphics[width = \linewidth]{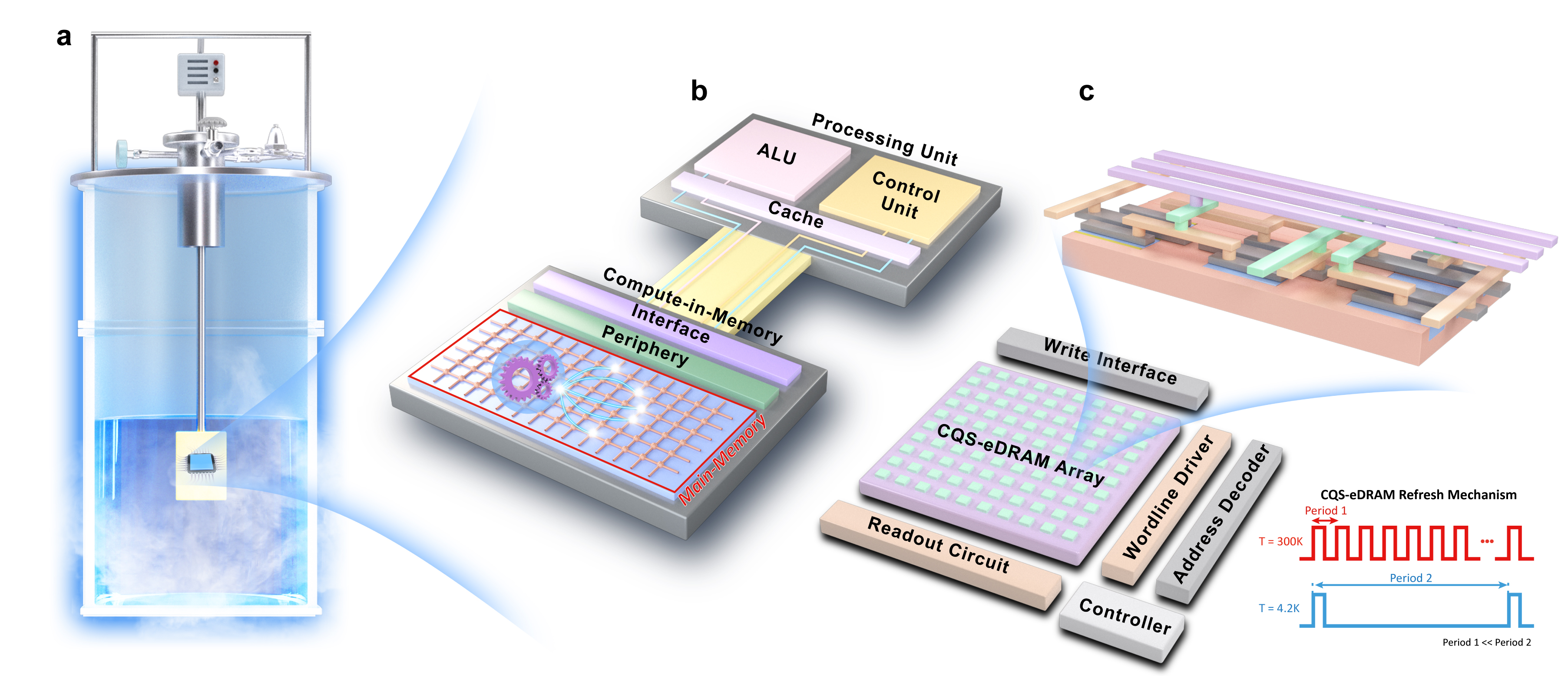}}
    \caption{
        \textbf{Architecture of CQS-eDRAM.} \textbf{a}, A cryogenic test setup with the test chip placed inside a dewar.
        \textbf{b}, The overall CIM-based computing architecture consists of a processing unit and a CIM module. The processing unit comprises an arithmetic logic unit (ALU), a controller, and an on-chip cache, while the CIM module for specialized algorithms includes interfaces, peripheral circuits, and main memory.
        \textbf{c}, The configuration of the proposed CQS-eDRAM array and its periphery read/write control circuits. This memory operates at a much longer refresh period compared to room temperature due to the negligible leakage.
   }
    \label{fig:overall}
\end{figure*}

As the logic-memory gap continuously enlarges over time, memory access has become the major bottleneck of computation performance in data-intensive applications \cite{cim_memory_wall,cim_energy_isscc2014,cim_memory_wall_ne_2018}. 
One promising solution is compute-in-memory (CIM), which is commonly adopted to alleviate the overhead between processing units and memory due to data transfer \cite{cim_review_nature_2018,cim_review_NE_2018,cim_review_CSM_2021}.
In general, for energy-efficient CIM implementation, the memory design needs to meet the high speed, high capacity, high reliability, and low power consumption requirements \cite{cim_review_nature_2020,cim_review_SSCM_2019,cim_challenge_vlsi_2019}. 
Figure~\ref{fig:overall} illustrates a typical CIM framework, where the memory hierarchical topology can be implemented with static random-access memory (SRAM), dynamic RAM (DRAM), or non-volatile memory modules like resistive RAM (RRAM) and magnetic RAM (MRAM) \cite{cim_sram_isscc_2020, cim_sram_7nm_isscc_2020,cim_sram_isscc_2021, cim_sram_isscc_2022, cim_rram_nature_2022,cim_rram_simpl_nature_2022,cim_rram_ne_2021,cim_rram_multi_bit_ne_2021, cim_mram_isscc_2023,cim_AeDRAM_isscc_2021,cim_eDRAM_isscc_2021,cim_eDRAM_Confvlsi_2022}.
Among the available memory technologies, embedded DRAM (eDRAM) stands out as an appealing candidate due to its process compatibility and high density \cite{room_edram_2t_jssc_2009, room_edram_3t_jssc_2011, room_edram_ifgc_jssc_2018, room_edram_ifgci_tcasi_2018,cryo_eDRAM_cryomem_Confcicc_2021,room_edram_3ttggc_tvlsi_2016}. 
However, unlike cross-coupled SRAM circuits, eDRAM lacks a latch design, which inevitably results in data floating.
Accordingly, a mandatory refresh operation (i.e., with the refresh period ranging from $\mu$s to ms) is introduced to maintain data reliability, which in turn leads to additional power consumption and reduced data access efficiency compared to SRAM \cite{dram_refresh_Confisca_2012,dram_refresh_TC_2016}.
As a result, the dynamic storage of eDRAM limits the computational energy efficiency in room-temperature CIM, especially for complex neural network computing applications \cite{cim_eDRAM_isscc_2021,cim_AeDRAM_isscc_2021,cim_eDRAM_Confvlsi_2022}.

Given that the inherent dynamic storage characteristic of eDRAM lies in the limited retention time (ranging from $\mu$s to ms) caused by leakage \cite{model_leakage_micro_2006,model_leakage_2_computer_2003}, previous efforts have aimed to optimize its performance through various measures \cite{room_edram_2t_jssc_2009, room_edram_3t_jssc_2011, room_edram_ifgc_jssc_2018, room_edram_ifgci_tcasi_2018,cryo_eDRAM_cryomem_Confcicc_2021,room_edram_3ttggc_tvlsi_2016}. For example, to extend the data retention time, internal feedback has been proposed to compensate for the leakage of the storage node \cite{room_edram_ifgc_jssc_2018,room_edram_ifgci_tcasi_2018}; meanwhile, wordline voltage boosting techniques have been adopted to ensure reliable storage of `1' or `0', yet such strategies suffer from the increased power consumption and deteriorated device reliability \cite{model_lifetime_TDMR_2021,model_lifetime2015_TNS_2015}.
Alternatively, based on the operating principle of metal-oxide-semiconductor field-effect transistor (MOSFET) (i.e., which is the building block of eDRAM), the leakage current $I_{\rm{Sub}}$ at the sub-threshold region exhibits an exponential relationship with temperature $T$ (i.e., $I_{\rm{Sub}}\propto{exp(-eV/kT)}$, where $e$ is the electron charge, $k$ is Boltzmann constant, and $V$ is the applied gate voltage of transistor). 
In this context, the low-leakage mode of MOSFETs at low temperatures can, in principle, significantly enhance the robustness of data storage in eDRAM cells without invoking the refresh operations. 
Accordingly, integrating this cryogenic quasi-static eDRAM (CQS-eDRAM) module into the CIM architecture (Fig.~\ref{fig:overall}b) would not only increase the storage density (i.e., owning to the simplified memory circuitry), but also improve the computational efficiency of the system.

\begin{figure*}[tb]
    \centerline{\includegraphics[width = \linewidth]{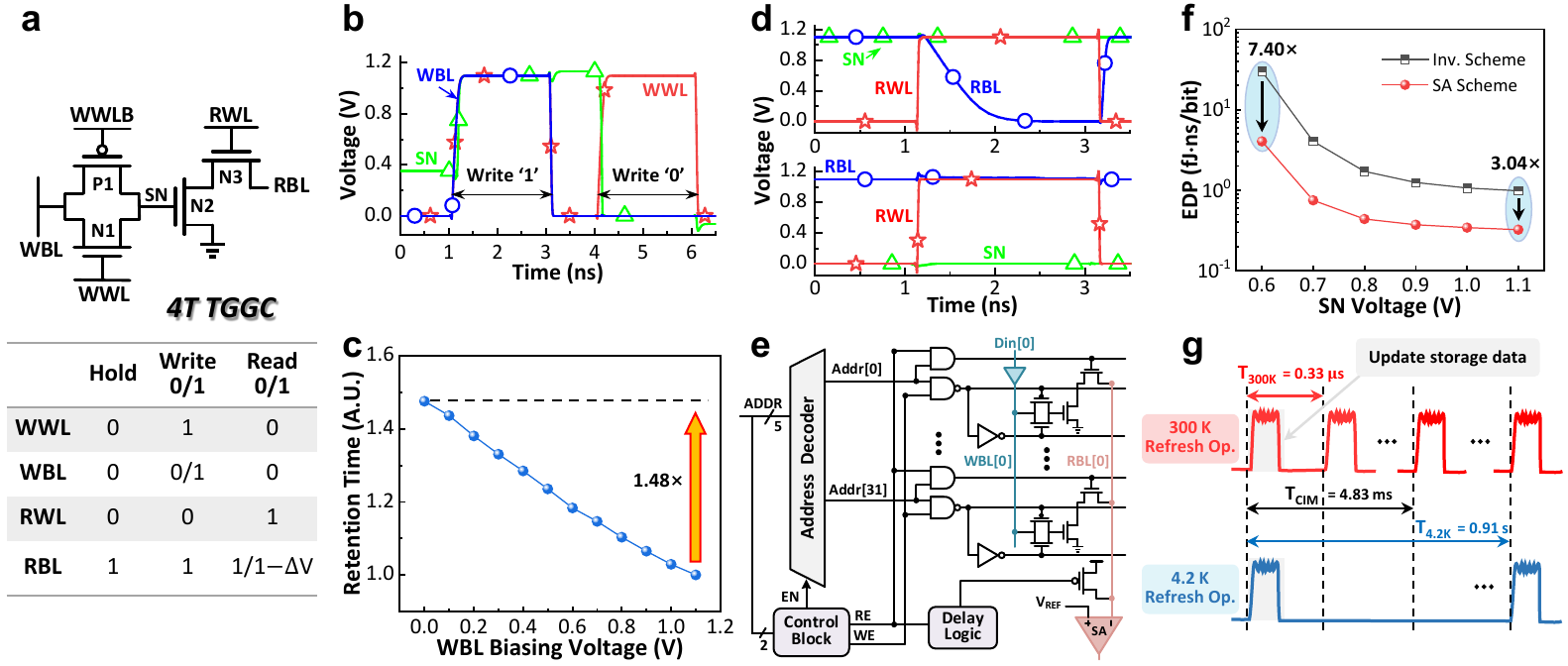}}
    \caption{
    \textbf{4T TGGC bit-cell design of the CQS-eDRAM.} \textbf{a}, The 4T TGGC schematic and the signal voltages in each operation. \textbf{b}, The waveform of 4T TGGC in a write operation, demonstrating the rail-to-rail full-swing data storage. \textbf{c}, Retention time versus different WBL biasing voltages. \textbf{d}, The waveform of 4T TGGC in read `1' (top panel) and `0' (bottom panel) operations with fast data access. \textbf{e}, Schematic block diagram of the 4-Kb CQS-eDRAM. \textbf{f}, The effect of SN voltage on read speed and power at cryogenic temperatures. \textbf{g}, Measured refresh operation waveforms (refresh signal) of the CQS-eDRAM array in reference to the CIM cycle ($T_{\rm{CIM}}$) at 300 K and 4.2 K.
    }
    \label{fig:bitcell}
\end{figure*}

Inspired by the aforementioned proposal, in this article, we present the utilization of CQS-eDRAM for energy-efficient CIM applications.  
Based on the accurate cryogenic device model and process design kit (PDK), we present a cryogenic four-transistor transmission gate gain-cell (4T TGGC) topology that enables quasi-static storage operation at cryogenic temperatures by fully leveraging its advantages in terms of reduced leakage and wire delay. 
Furthermore, we employ the cryogenic write bitline (WBL) biasing technique and dedicated readout circuitry to optimize the power consumption of the read and write operation.
Experimental data of the 4Kb CQS-eDRAM chip reveals a remarkable retention time of 66.50 seconds, with a more uniform retention time distribution at $T$ = 4.2 K. The quasi-static storage operation contributes to an overall energy efficiency improvement of 2.63 times when executing the ResNet-18 network.
In addition, with the utilization of dynamic voltage scaling (DVS) and dynamic refresh period scaling (DRPS) techniques, CQS-eDRAM achieves a 7.1\% decrease in retention power and a 13.6\% reduction in dynamic power under a tolerable data error rate.

\section*{Constructing cryogenic quasi-static 4T TGGC cell}
In our previous study, we conducted a comprehensive temperature-dependent characterization of NMOS and PMOS devices in Huali Microelectronics Corporation (HLMC) 40-nm low-power (40LP) technology, and revealed the underlying mechanisms of key electrical parameters, such as threshold voltage ($V_{\rm{th}}$), on-state current ($I_{\rm{on}}$), and transconductance ($g_{\rm{m}}$), at cryogenic temperatures \cite{shiming_book,model_zewei_device_EDL_2020,model_zewei_circuit_ConfEDTM_2021,model_zhidong_JEDS_2022}. 
Based on the device's physical models, we also developed a modified BSIM compact model and a generic PDK that can be applied to full-size devices across the entire temperature range, making cryogenic complementary metal–oxide–semiconductor (CMOS) based VLSI design possible \cite{model_zewei_cryopdk_arxiv_2022} (Supplementary Fig. S1).

Benefiting from this platform, we managed to design a suitable eDRAM architecture for low-temperature operation.
Firstly, we compared the write performance of different eDRAM bit-cell designs (from 2T to 4T \cite{cryo_eDRAM_cryomem_Confcicc_2021,room_edram_2t_jssc_2009, room_edram_3t_jssc_2011, room_edram_ifgc_jssc_2018,room_edram_ifgci_tcasi_2018, room_edram_3ttggc_tvlsi_2016}) at cryogenic temperatures. 
It is seen that a single-type write port controlled by the write wordline (WWL) is less effective due to the degraded signal written into the storage node (SN), as shown in Supplementary Figs. S2a-c. 
To address this issue, the wordline voltage boosting technique has been widely used to ensure a significant initial voltage difference by over-driving the gate voltage.
However, at cryogenic temperatures, this strategy becomes less attractive due to the $V_{\rm{th}}$ shift of $\Delta$$V_{\rm{th}}>$ 0.11 V (Supplementary Fig. S1b), which in turn brings about more power and performance loss for single-type write port designs \cite{cryo_eDRAM_cryomem_Confcicc_2021,room_edram_2t_jssc_2009, room_edram_3t_jssc_2011, room_edram_ifgc_jssc_2018, room_edram_ifgci_tcasi_2018}, as shown in Supplementary Figs. S2f-g. 
Alternatively, in our design, we adopted the transmission gate (TG) based write port topology (Fig.~\ref{fig:bitcell}a), where the parallel-configured NMOS and PMOS pair (controlled by WWL and WWLB) warrants the rail-to-rail swing during the write operation, as shown in Fig.~\ref{fig:bitcell}b.
Quantitatively, our simulation results demonstrate that our 4T TGGC bit-cell achieves a 1.84$\times$ (1.61$\times$) improvement in the energy consumption of write operation compared to a single N-type (P-type) write port at 10 K (Supplementary Fig. S2f). 
Concurrently, the TG port also obtains a 2.10$\times$ (1.76$\times$) improvement in the write access time, as shown in Supplementary Fig. S2g.
Moreover, it is found that the WBL biasing voltage during non-write access periods plays a crucial role in determining the retention time of GC-eDRAM designs.
Figure~\ref{fig:bitcell}c visualizes the negative correlation between the retention time and WBL biasing voltage, and the optimal biasing condition of the WBL at cryogenic temperatures is found to be 0 V, which manages a 1.48$\times$ improvement in retention time compared to that at V = $V_{\rm{DD}}$.

In view of the read operation, conventional eDRAM bit-cells typically use a 1T read port to save the layout area \cite{cryo_eDRAM_cryomem_Confcicc_2021, room_edram_ifgc_jssc_2018, room_edram_ifgci_tcasi_2018,room_edram_3ttggc_tvlsi_2016}.
However, the unselected read wordline (RWL) will impede the read performance, resulting in longer access times and more power overhead.
To achieve non-destructive and high-speed read operations, we exploited the 2T-NMOS read port (Fig.~\ref{fig:bitcell}a), as the driving strength of the NMOS transistor surpasses that of the PMOS counterpart at cryogenic temperatures \cite{model_zewei_device_EDL_2020,model_zewei_circuit_ConfEDTM_2021,model_zhidong_JEDS_2022}.
The simulated read waveforms in Fig.~\ref{fig:bitcell}d confirm successful read `0' and `1' operations, where the read bitline (RBL) only discharges during the read `1' operation.
In accordance with our simulation results, our proposed 4T TGGC eDRAM achieves a 1.98$\times$ improvement in energy overhead and a 1.41$\times$ improvement in read access time compared to the 1T-NMOS read port configuration, as shown in Supplementary Figs. S2f-g.
In addition to the single bit-cell optimization, we also redesigned the readout circuit of the 4Kb CQS-eDRAM for cryogenic operation with the consideration of speed and energy consumption. 
As illustrated in Fig.~\ref{fig:bitcell}e, the sense amplifier (SA) with an additional reference voltage ($V_{\rm{REF}}$) was employed in the readout circuit, which is divided into 32 rows and 128 columns. 
Figure~\ref{fig:bitcell}f presents the simulation results of the energy-delay-product (EDP) versus the SN voltage for differential SA and inverter (Inv) configurations. 
By comparing these two readout circuit structures, we can observe that the SA structure consistently exhibits a smaller EDP than the Inv counterpart, regardless of the SN voltage. 
As the SN voltage decreases from $V_{\rm{DD}}$ to 0.6 V, the difference in EDP between the two readout circuits further increases (from 3.04$\times$ to 7.40$\times$). 

Consistent with the simulation predictions, once data are written into the bit-cell, their information can be well-retained even after 188 CIM clock cycle times (approximately 0.91 s) without invoking the refresh operation, therefore validating the quasi-static mode at cryogenic temperatures (Fig.~\ref{fig:bitcell}g).
Moreover, it should be highlighted that our 4T TGGC bit-cell offers a layout area reduction of up to 23\% compared to the conventional 6T-SRAM with equivalent intrinsic delay, thus providing greater storage capacity when implemented at the cache level of the cryogenic CIM system.

\section*{Retention time and power characterizations of the 4Kb CQS-eDRAM chip}
\label{section: rt}

\begin{figure*}[tb]
    \centerline{\includegraphics[width = 1\linewidth]{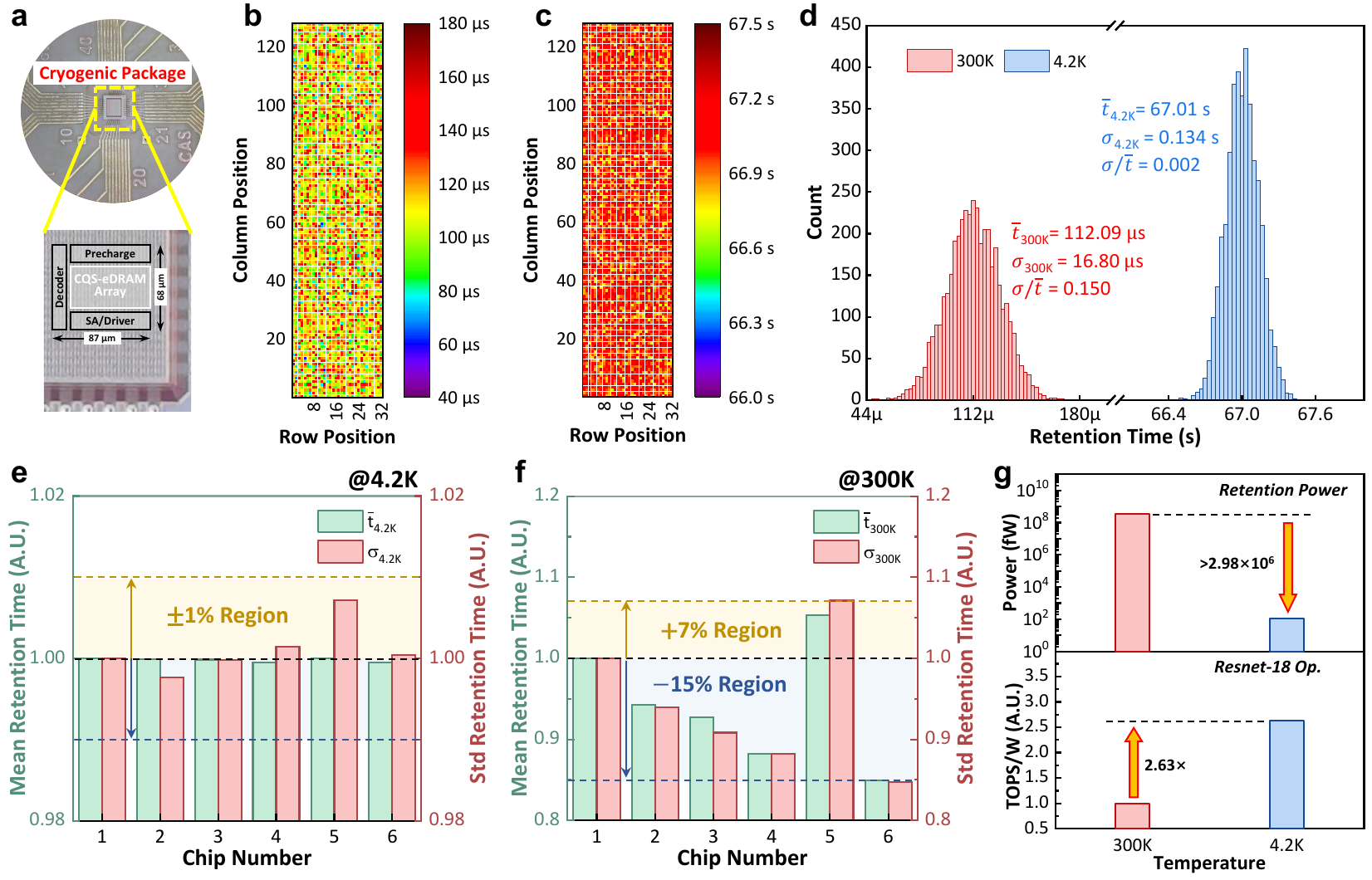}}
    \caption{
    \textbf{Retention time and power characterizations of the 4Kb CQS-eDRAM Chip.} \textbf{a}, Cryogenic package and the die-photo of the CQS-eDRAM chip. Retention heat maps of the 4Kb CQS-eDRAM array (Chip 1) at \textbf{(b)} 300 K and \textbf{(c)} 4.2 K. \textbf{d}, The corresponding retention time histogram manifests the normal distributions of retention time at 300 K (red) and 4.2 K (blue). The statistical retention $\overline{t}$ and $\sigma$ variations across Chip 1 to Chip 6 at \textbf{(e)} 4.2 K and \textbf{(f)} 300 K, where values are normalized relative to Chip 1, and the mean and standard deviation of Chip 1 are defined as $\overline{t}_{\rm{0}}$ and $\sigma_{\rm{0}}$, respectively. \textbf{g}, The significant improvement in retention power consumption and the ResNet-18 network computation efficiency of the 4Kb CQS-eDRAM at 4.2 K.
    }
    \label{fig:rt_single}
\end{figure*} 

Following the optimized 4T-TGGC cell design, we designed the 4Kb CQS-eDRAM chip using the 40LP process. 
Figure~\ref{fig:rt_single}a shows the cryogenic package and the die-photo of our 4Kb CQS-eDRAM. 
A total of six test chips labeled as Chip 1 to Chip 6 were prepared.
For cryogenic measurement, the I/O pads of the chip were directly wired-bonded to the cryogenic package board, as shown in Supplementary Fig. S3. 
Afterwards, the test chip was connected to a field-programmable gate array (FPGA) board through a customized conversion printed circuit board (PCB) to enable the control signal transfer and data processing. The sampled data on the FPGA was then sent to the host PC for subsequent data processing.

Figures ~\ref{fig:rt_single}b-c display the quantitative retention time heat mappings of Chip 1 at 300 K and 4.2 K, respectively.
Strikingly, a remarkable increase in the mean retention time ($\overline{t}$) is observed, increasing from 112.09 $\mu$s to 67.01 s as the base temperature drops to 4.2 K, as shown in Fig.~\ref{fig:rt_single}d.
Such a significant improvement (i.e., which is six orders of magnitude greater than that at 300 K) can be mainly owing to the suppression of sub-threshold current and reverse-biased junction diode leakage, both of which are exponentially dependent on the temperature \cite{model_leakage_micro_2006, model_leakage_2_computer_2003}.
Likewise, the standard deviations (std, $\sigma$) of the retention time across the entire memory array are around 16.80 $\mu$s at 300 K and 134 ms at 4.2 K.

\begin{figure*}[tb]
    \centerline{\includegraphics[width = \linewidth]{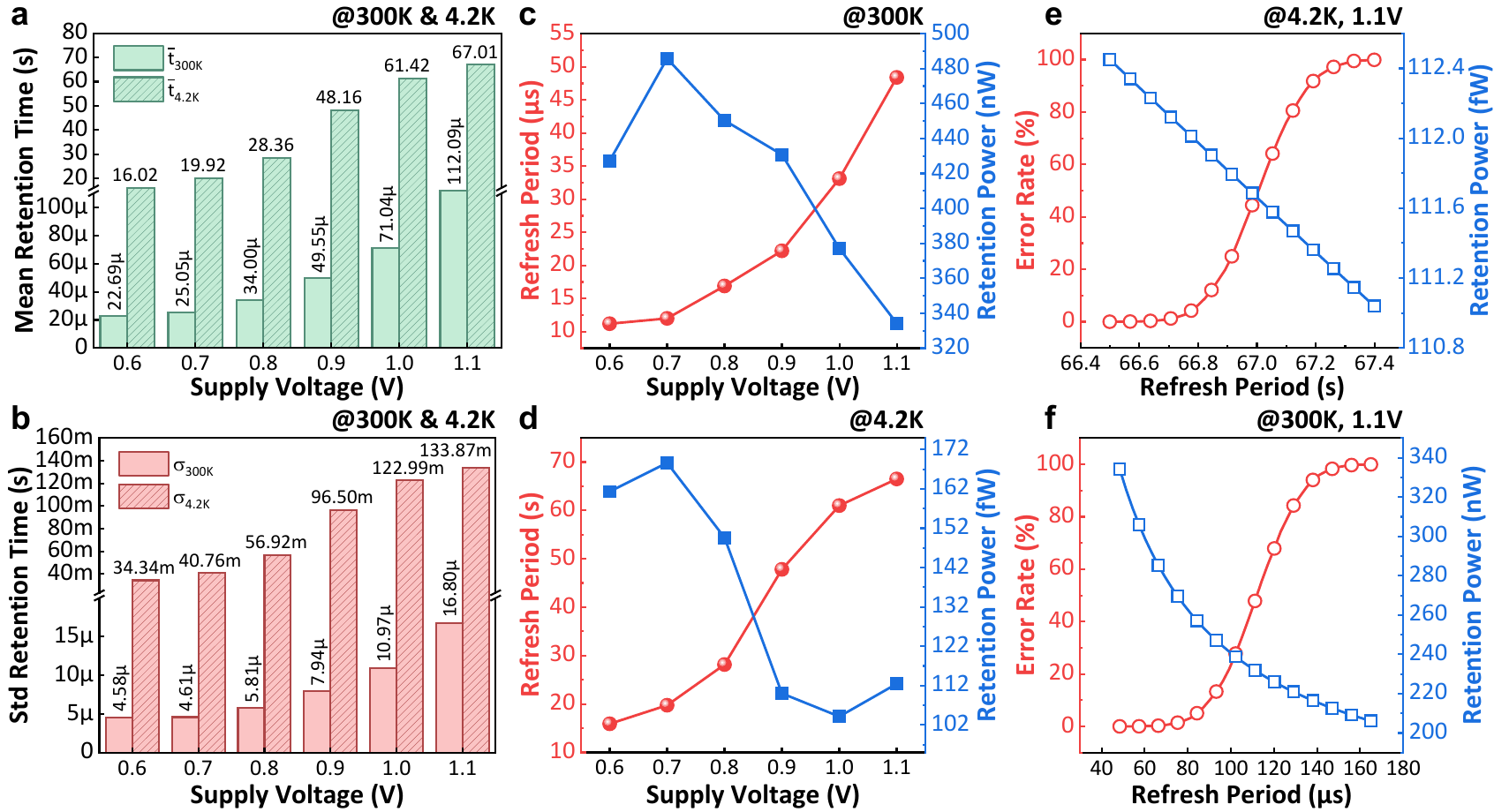}}
    \caption{
    \textbf{The effect of DVS and DRPS on CQS-eDRAM performance.} The comparison of \textbf{(a)} mean and \textbf{(b)} standard deviation of retention time with respect to the supply voltage at 4.2 K and 300 K. Supply voltage-dependent refresh period and the array retention power at \textbf{(c)} 300 K and \textbf{(d)} 4.2 K. 
    The error rate and the retention power of 4Kb CQS-eDRAM array versus different refresh periods at \textbf{(e)} 4.2 K and \textbf{(f)} 300 K. 
    }
    \label{fig:analysis}
\end{figure*}

Apart from $\overline{t}$, retention time variation serves as another crucial parameter for benchmarking the eDRAM performance. 
Accordingly, the normalized mean (green bars) and standard deviation (red bars) retention times of the six CQS-eDRAM chips are summarized in Figs.~\ref{fig:rt_single}e-f.
Benefiting from the diminished leakage current and thermal noise, the normalized retention time remains nearly constant at $T$ = 4.2 K, with $\overline{t}_{\rm{4.2 K}}/\overline{t}_{\rm{0}}\approx1\pm0.01$. The $\sigma_{\rm{4.2 K}}/\sigma_{\rm{0}}$ values also exhibit a negligible chip-to-chip variation, as shown in Fig.~\ref{fig:rt_single}e.
In contrast, when the CQS-eDRAM chips warm up to $T$ = 300 K, the increased thermal noise activation would introduce more current fluctuations.
As a result, both the measured $\overline{t}_{300 K}$ and $\sigma_{300 K}$ vary from -15\% to +7\% across the six chips (Fig.~\ref{fig:rt_single}f).
To evaluate the dispersion of retention time at different temperatures, we adopted the $\sigma/\overline{t}$ as the figure-of-merit, which enables fair comparisons among the datasets with different scales.
As highlighted in Fig.~\ref{fig:rt_single}d, the $\sigma/\overline{t}$ value decreases from 0.150 at 300 K to 0.002 at 4.2 K, which is equivalent to a 75$\times$ improvement in stability.

Based on the above characteristic pros, the quasi-static storage operation and dramatic enhancement of retention time enable the integration of our 4Kb CQS-eDRAM into the CIM framework, replacing the conventional 6T-SRAM module adopted at room temperature. 
To justify this proposal, we conducted a comprehensive evaluation by employing the CQS-eDRAM as the main memory in CIM for a complete ResNet-18 network (Supplementary Fig. S4).
As shown in Supplementary Figs. S4f-h, the energy efficiency reaches 1232.74 TOPS/W at $T$ = 4.2 K, which equals a 2.63$\times$ increase compared to 300 K. Furthermore, this significant enhancement is accompanied by a substantial 1.94$\times$ increase in the processing speed (i.e., in terms of frame rate) at 4.2 K (Supplementary Fig. S4f). In addition, when performing the same training task within an identical time frame, the energy consumption at 300 K is 1.36$\times$ higher than that at 4.2 K, as shown in Supplementary Fig. S4g. 
It is worth mentioning that a significant 62\% of the total energy is allocated to refresh operations during the ResNet-18 network computation at room temperatures. 
In contrast, the energy allocated to refresh operations during the ResNet-18 network computation is negligible due to the quasi-static storage mode at cryogenic temperatures.
These findings clearly demonstrate that the cryogenic CIM equipped with CQS-eDRAM can empower nearly all energy for computational tasks.
In view of the overall performance, Fig.~\ref{fig:rt_single}g unveils that our CQS-eDRAM yields a 2.98$\times$10$^6$ times improvement in retention power, while achieving an energy efficiency gain of nearly 2.63 times at $T$ = 4.2 K compared to room-temperature data. 
Consequently, our results highlight the potential of CQS-eDRAM for high-performance computation.

\section*{Optimization of CQS-eDRAM performance for low-power applications} 
Considering both the retention time and error rate of the eDRAM array rely on the supply voltage and refresh period, we are able to utilize DVS and DRPS strategies to further optimize the CQS-eDRAM performance \cite{dram_dvfs_ConfDAC_2018,dram_dvfs_ConfAC_2011,dram_dvfs_ConfISLPED_2013}.
Figures~\ref{fig:analysis}a-b show the mean and standard deviation of the retention time for Chip 1 with the supply voltage $V_{\rm{DD}}$ ranging from 0.6 V to 1.1 V at 4.2 K and 300 K, respectively. 
It is observed that both $\overline{t}$ and $\sigma$ exhibit a positive correlation with $V_{\rm{DD}}$: $\overline{t}_{\rm{4.2 K}}$ ($\overline{t}_{\rm{300 K}}$) is elevated from 16.02 s (22.69 $\mu$s) at 0.6 V to 67.01 s (112.09 $\mu$s) at 1.1 V, and the corresponding $\sigma_{\rm{4.2 K}}$ ($\sigma_{\rm{300 K}}$) obtains a 3.90$\times$ (3.67$\times$) improvement.

\begin{table*}[t]
\caption{\centering{Comparison with the different cryogenic memory designs for high-performance computation.}}
\vspace{-1.2em}
\begin{center}
\begin{threeparttable}
\resizebox{0.96\textwidth}{!}{
\begin{tabular}{|l|c|c|c|c|c|c|c|}
\hline
\rule{0pt}{9pt}
& Duzer \cite{cryo_sup_JJC_TAS_2013}
& Tanaka \cite{cryo_sup_JJNTC_TAS_2016}
& Lee \cite{cryo_dram_Confisca_2021}
& Saligram \cite{cryo_eDRAM_cryomem_Confcicc_2021}
& \textbf{This Work}  \\ \hline

\rule{0pt}{9pt}
Technology
&  JJ+CMOS
&  JJ+nTron+CMOS
&  1T1C-DRAM
&  2T-eDRAM
&  4T-eDRAM       \\ \hline

\rule{0pt}{9pt}
Si Integration
& \multicolumn{2}{c|}{\begin{tabular}[c]{@{}c@{}} Heterogeneous Integration \end{tabular}}
& \multicolumn{3}{c|}{\begin{tabular}[c]{@{}c@{}} Monolithic Integration       \end{tabular}}
 \\ \hline

\rule{0pt}{9pt}
Temperature
& \multicolumn{2}{c|}{\begin{tabular}[c]{@{}c@{}} 4.2K \end{tabular}}
& \multicolumn{1}{c|}{\begin{tabular}[c]{@{}c@{}} 77K - 300K \end{tabular}}
& \multicolumn{2}{c|}{\begin{tabular}[c]{@{}c@{}} 4.2K - 300K  \end{tabular}}
 \\ \hline

\rule{0pt}{9pt}
Process Node
& 65 nm
& 65 nm
& N/R
& 28 nm
& 40 nm      \\ \hline

\rule{0pt}{9pt}
Supply Voltage
& 1 V
& N/R
& 1.2 V
& 0.9 V
& 1.1 V       \\ \hline

\rule{0pt}{9pt}
Memory Cell
& SRAM
& DRAM
& DRAM
& eDRAM
& eDRAM       \\ \hline

\rule{0pt}{9pt}
Cell Area
& 3.36 $\mu$m$^2$
& N/R
& N/R
& 0.57 $\mu$m$^2$
& 0.7 $\mu$m$^2$      \\ \hline

\rule{0pt}{13pt}
Access Time
& {\begin{tabular}[c]{@{}c@{}} 430 ps Read \\ 300 ps Write                  \end{tabular}}  
& {\begin{tabular}[c]{@{}c@{}} 660 ps Read                  \end{tabular}}  
& 39.43 ns
& 763 ps
& \textbf{{\begin{tabular}[c]{@{}c@{}} 1100 ps (300K) \\ 820 ps (4.2K)  \end{tabular}}}       \\ \hline

\rule{0pt}{13pt}
Dynamic Power    
& {\begin{tabular}[c]{@{}c@{}} 12 mW Read \\ 21 mW Write                  \end{tabular}}  
& {\begin{tabular}[c]{@{}c@{}} 0.78 mW Read \\ 2.2 mW Write               \end{tabular}}
& N/R
& {\begin{tabular}[c]{@{}c@{}} 0.76 mW (300K)  \\ 0.56 mW (6K)            \end{tabular}}    
& \textbf{{\begin{tabular}[c]{@{}c@{}} 131 $\mu$W (300K) \\ 108 $\mu$W (4.2K)  \end{tabular}}}       \\ \hline

\rule{0pt}{13pt}
Retention Time
& N/A
& N/A
& {\begin{tabular}[c]{@{}c@{}} 2.4 ms (300K) \\ 1.28 s (77K) \end{tabular}}
& {\begin{tabular}[c]{@{}c@{}} 2.4 $\mu$s (300K) \\ 6.5 s (4.2K)            \end{tabular}}
& \textbf{{\begin{tabular}[c]{@{}c@{}} 48.40 $\mu$s (300K) \\ 66.50 s (4.2K)  \end{tabular}}}       \\ \hline

\rule{0pt}{13pt}
{\begin{tabular}[c]{@{}c@{}} Retention Power \\ per Kb            \end{tabular}} 
& N/A
& N/A
& $>$19.22$^*$ pW (77K)
& {\begin{tabular}[c]{@{}c@{}} 350$^*$ nW (300K) \\ 118.15$^*$  fW (4.2K)            \end{tabular}}
& \textbf{{\begin{tabular}[c]{@{}c@{}} 83.56$^\dagger$ nW (300K) \\ 28$^\dagger$ fW (4.2K)  \end{tabular}}}       \\ \hline

\end{tabular}
}
    \begin{tablenotes}
        \footnotesize             
        \item JJ: Josephson Junction\quad nTron: Nanacryotrons\quad N/A: Not Applicable\quad N/R: Not Reported
        \item $^*$Calculated from the reported data 
        \item $^\dagger$Off-chip and IO power are not included
    \end{tablenotes}

\end{threeparttable}
\end{center}
\label{table:1}
\end{table*}

In view of the impact of $V_{\rm{DD}}$ on CQS-eDRAM operation, Figs.~\ref{fig:analysis}c-d display the evolutions of the refresh period $t_{\rm{min}}$ (i.e., which is defined as the minimum time duration to ensure 100\% yield of the stored data) and retention power (i.e., $P_{\rm{retention}}=(E_{\rm{read}}+E_{\rm{write}}+E_{\rm{leakage}})/t_{\rm{min}}$, where $E_{\rm{read}}$ and $E_{\rm{write}}$ are the total energy consumed during read and write operation for the storage data, and $E_{\rm{leakage}}$ is the total leakage energy dissipated during the refresh process) as functions of supply voltage for the CQS-eDRAM at 300 K and 4.2 K, respectively. 
For the room-temperature case (Fig.~\ref{fig:analysis}c), it is seen that a higher $V_{\rm{DD}}$ helps to extend the retention time, and the optimization of retention power through voltage scaling reveals that the minimum retention power is attained at 1.1 V, yet it comes at the expense of the highest dynamic power consumption of 131 $\mu$W.
On the contrary, when the CQS-eDRAM is operated at $T$ = 4.2 K (Fig.~\ref{fig:analysis}d), $V_{\rm{DD}}$ = 1.0 V is identified as the optimal operating condition for realizing both the minimum retention power of 104 fW (a 7.1\% reduction compared to the case of $V_{\rm{DD}}$ = 1.1 V) and resulting in a 13.6\% reduction in dynamic power consumption. 

Besides DVS, DRPS is another useful tuning method for eDRAM optimization \cite{dram_drps_ConfHPCA_2014, dram_drps_ConfICCAD_2020}. 
According to the operating principle of eDRAM, the increase in the refresh period leads to a higher error rate in memory operations \cite{cryo_dram_Confisca_2021,room_edram_ifgci_tcasi_2018}. 
This observation is consistent with our experimental results at $T$ = 4.2 K and 300 K, as shown in Figs.~\ref{fig:analysis}e-f. 
Notably, the error rate at 4.2 K is more sensitive to the refresh period compared to the data obtained at room temperature, possibly attributed to the fact that the refresh period becomes more tightly constrained, where even slight variations can introduce significant changes on the error rate.
In practical applications, it is crucial to ensure that the refresh periods remain shorter than 66.50 s to mitigate the error rate issue.
Based on the supply-voltage dependent data presented in Fig.~\ref{fig:analysis}, it is advisable for energy-efficient applications to adopt a combined strategy of DVS and DRPS to meet the low power budget requirements. On the other hand, for applications that prioritize computational accuracy and performance, DVS proves to be a valuable approach for reducing the system-level power consumption while maintaining the desired performance benchmark.

\section*{Conclusions}
In conclusion, we have presented a 4T TGGC eDRAM bit-cell design with quasi-static memory operation mode at cryogenic temperatures.
From the low-temperature measurements, we have demonstrated that the TG-based write port can achieve high-quality write operation without the need for a wordline boosting technique, and the 2T-NMOS read port enables faster and more energy-efficient operations.
Implemented in a 40LP process, our 4Kb CQS-eDRAM chip achieves a retention time of 66.50 s at 4.2 K (1.37$\times$10$^6$ times higher than that at 300 K), with an array retention power of 112 fW (i.e., 28 fW/Kb) for 100\% yield. Besides, it also outperforms other designs at cryogenic temperatures in terms of retention time, dynamic power, and retention power, as summarized in Table~\ref{table:1}~\cite{cryo_sup_JJC_TAS_2013, cryo_sup_JJNTC_TAS_2016, cryo_dram_Confisca_2021, cryo_eDRAM_cryomem_Confcicc_2021}.
Moreover, our 4T TGGC-based CQS-eDRAM design provides significant reductions in both dynamic and retention power, along with a more compact bit-cell area compared to 6T SRAM, making it an appealing candidate for high-density and low-power memory implementation for cryogenic computing applications.

\bibliographystyle{unsrt}
\bibliography{cryo}

\section*{Acknowledgements} 
This work was supported in part by the National Key R\&D Program of China under contract no. 2021YFA0715503, in part by the National Natural Science Foundation of China (grant nos. 62074101, 92164104, 62150710549, 62220106011), in part by the Shanghai Science and Technology Commission Funding under Grant 20ZR1435800, and in part by the Zhangjiang Laboratory. X.-F.K. acknowledges support from the Strategic Priority Research Program of CAS (XDA18010000), and the Shanghai Rising-Star Program (grant no. 21QA1406000).

\section*{Author contributions}
Y.-H.S. and Y.-J.H. conceived the concept of this work. 
Y.-H.S. and H.-T.Z. designed the memory chip.
H.S., M.-R.Z., and Q.D. built the test platform and finished the cryogenic experiments measurement. 
Z.-D.T., Y.-M.Y., Y.-Q.H., Y.G., and X.-F.K. contributed to the cryogenic device model calibration. 
Y.-H.S., H.-T.Z., X.-F.K., M.-R.Z., W.-F.Z., and Y.-J.H. wrote the manuscript. 
X.-F.K. and Y.-J.H. supervised this work. 
All the authors contributed to the results and discussion and commented on the manuscript.

\balance

\end{document}